\documentclass[10 pt, letter]{article}
\usepackage{graphicx}
\usepackage{caption}
\usepackage{amsmath}
\usepackage{amsthm}
\usepackage{enumerate}
\usepackage{subfig}
\usepackage{float}
\usepackage{amsfonts}
\usepackage{amssymb}
\usepackage{anysize}
\usepackage{helvet}
\usepackage{setspace}
\usepackage{multirow}
\usepackage{tabularx}
\usepackage{tabulary}
\usepackage{longtable}
\usepackage{float}
\usepackage{mathtools}
\usepackage[none]{hyphenat}
\title{\bf 
Analytical  model for the dynamical motion of the bulges  of two interacting galaxies}
\author{Elkin L. L\'opez\footnote{lopezlopezelkin@gmail.com}, Gustavo V. L\'opez\footnote{gulopez@cencar.udg.mx}\\Ê\\
 Departamento de F\'{i}sica, Universidad de Guadalajara,\\
 Blvd. Marcelino Garc\'{i}a Barragan y Calzada Ol\'{i}mpica, \\ CP 44200, Guadalajara, Jalisco, Mexico, \\ \\
 Simon N. Kemp
\footnote{snk@astro.iam.udg.mx}~\\ \\
Instituto de Astronomia y Meteorologia, Universidad de Guadalajara,\\ 
Av Vallarta 2602, Col Arcos Vallarta, \\
CP 44130, Guadalajara, Jalisco, Mexico,  
 }
\begin{document}
\maketitle

%\centerline{\large\bf Abstract}
\begin{abstract}
\noindent
Two mathematical models of three bodies of variable masses are used to obtain a qualitative  description of two interacting galaxies  with mass exchange and mass loss. The reference system  is centred on the largest body (the most massive  galaxy), and the other two bodies are allowed to move around  this one under the  laws of gravity. The third body, which simulated the mass lost by the second galaxy in the form of a tail, increases its mass due to the mass lost by the second body and follows its trajectory.  We are interested in knowing the time evolution of the  separation of the two bulges of the interacting galaxies, and the parameters for the analytical models are obtained by running simulations with the GADGET-2 N-body  code. The resulting behavior of this distance in our mathematical models is qualitatively in good agreement with that obtained by this code.     
\end{abstract}
\vskip2pc\noindent
{\bf Keywords:} galaxies: kinematics and dynamics, galaxies: interactions, galaxies: bulges, galaxies: binary
\vskip1pc\noindent
{\bf PACS:} 98.35.Gi, 98.35.Jk, 9865.At, 98.65.Fz
\newpage
\section{ Introduction}
Galaxies are one of the most important structures in our Universe, and studies of their formation and subsequent evolution continue to fill the research journals \cite{white,dekel}.%1
~All types of galaxy consist of the combination of an elliptical-like component (bulge) \cite{bouok,milion,bertin,launhardt,dwek,xu,salviander}%2
~and a disk-like component \cite{vanture,bertin,giallongo,kennicutt,kenni2,burkert}, in different proportions (which can be zero),%3
~containing stars, gas, dust  \cite{bertin,kacprzak,ivison,daylan},%4
~plus a dark matter halo which usually contains the bulk of the mass. Visually the bulge component \cite{holtzman,magorrian}
 defines the trajectory of the galaxy, usually coinciding with the centre of the dark matter halo.%5

~ Galaxies exist in binary systems and in small groups (less than 100 members) or large clusters of galaxies with increasingly complicated dynamics \cite{turnary,pauling,voit, gallaghere}.%7
~When two galaxies interact, there is an exchange of mass \cite{gutowski},%8
~and usually one (or both) of the galaxies lose  mass in the form of one \cite{kemp}, or more usually, two tidal tails \cite{anva,vaan,Kara}.%9
~Material in tidal tails may escape the system or fall back into the galaxy eventually \cite{hibbard, kemp}, or may be incorporated in tidal dwarf galaxies,
galaxies newly formed from material in tidal tails \cite{duc,kemp}.%10
%11.
The distribution of material around interacting galaxies and the forms of the galaxies themselves have been successfully modelled
using numerical N-body codes since the works of \cite{toomre}.

Analytically this phenomenon corresponds to a typical mass variation problem, which is called the Gyld{\'e}n-Mestschersky problem in astronomy 
\cite{Gyl,Mes1,Mes2,Lov,Bek2,Bek3}, %12
~which has had a long standing challenge for a complete analytical description \cite{Ber, Bek1, spivak, Somm,Loca}.

There have been some works \cite{soa} with a semi-analytical description of the dynamics of two point galaxies  {san1,san2,bleh1,bitre,ezo}  but as far as we know, there have been few attempts to consider the mass variation problem in binary galaxies \cite{Jean1,Jean2}. The analytical treatment of two interacting galaxies as a two-body problem cannot be considered satisfactorily since one needs to consider the dynamics of binary galaxies with mass loss or mass exchange between them.\\%13
\\
Based on the observations made on previous studies \cite{spivak} and \cite{lopez}, we consider that Newton's equations of motion for mass variation problems are still valid. In addition, we consider that the trajectory of a galaxy is defined mainly by its bulge (centre of the dark halo), and this is true even when two galaxies are interacting. We also consider that these bulges are treated as point bodies with variable mass, having one galaxy more massive than the other one. The less massive body loses mass (galaxy tails) which is going to be absorbed by a third body which also moves around the binary system, and it only increases its mass due to the mass loss from the second body. The parameters (mass variations) selected for our three bodies in the mass variation models  will be taken from  the N-body simulation program GADGET-2 \cite{gadget2,gadget2-2,gadget2-3} since the real evolution times of such systems are in Gyr.  The principal point is to see if this very simple model of three bodies (galaxies) could qualitatively say something  about the behavior of the  very complicated real motion of two interacting galaxies, using parameters from the N-body simulations.
The third body simulates a tidal tail, and if a second tail is present, it is effectively included in the mass of the first (most massive) body.
\section{Analytical Models}   
Consider three bodies of initial masses $m_1(0)=m_{10}$, $m_2(0)=m_{20}$, and $m_3(0)=m_{30}$ under their mutual gravitational attraction. Then, the equations of motion of these bodies from an arbitrary inertial  frame of reference are
\begin{subequations}
\begin{eqnarray}  
& &\frac{d(m_1{\bf v}_1)}{dt}=-G\frac{m_1m_2}{R_{12}^2}\widehat{R}_{12}-G\frac{m_1m_3}{R_{13}^2}\widehat{R}_{13}\label{e1},\\ \nonumber\\
& &\frac{d(m_2{\bf v}_2)}{dt}=+G\frac{m_2m_1}{R_{12}^2}\widehat{R}_{12}-G\frac{m_2m_3}{R_{23}^2}\widehat{R}_{23}\label{e2},\\�\nonumber\\
& &\frac{d(m_3{\bf v}_3)}{dt}=+G\frac{m_3m_1}{R_{13}^2}\widehat{R}_{13}+G\frac{m_3m_2}{R_{23}^2}\widehat{R}_{23}\label{e3},
\end{eqnarray}
\end{subequations}
where $G$ is the gravitational constant ($G=6.674\times 10^{-11}$ m$^3$kg$^{-1}$s$^{-2}$, \cite{rosetion})
${\bf v}_i=d{\bf x}_i/dt$ for $i=1,2,3$ represent the velocities of the bodies, being ${\bf x}_i$ i=1,2,3 their positions, and $\widehat{R}_{ij}$ for $i\not=j$ are the unitary vectors 
$\widehat{R}_{ij}={\bf R}_{ij}/R_{ij}$, with $R_{ij}=|{\bf R}_{ij}|$, being ${\bf R}_{ij}={\bf x}_i-{\bf x}_j$.
\subsection{Model-1}
Moving our reference system to the body with mass $m_1$ (${\bf x}_1={\bf 0}$), the first equation is not considered, and defining ${\bf x}_2={\bf x}$ and ${\bf x}_3={\bf x}'$, the following equations result,
\begin{subequations}
\begin{eqnarray}  
& &\frac{d(m_2{\bf v})}{dt}=-G\frac{m_2m_1}{|{\bf x}|^{3/2}}{\bf x}+G\frac{m_2m_3}{|{\bf x'-x}|^{3/2}}({\bf x'}-{\bf x})\label{em11},\\�\nonumber\\
& &\frac{d(m_3{\bf v}')}{dt}=-G\frac{m_3m_2}{|{\bf x'-x}|^{3/2}}({\bf x'-x})-G\frac{m_3m_1}{|{\bf x'}|^{3/2}}{\bf x'}\label{em12},
\end{eqnarray}
\end{subequations}
where the time depending mass will be  taken as
\begin{subequations}
\begin{eqnarray}
& &m_1(t)=m_{10}+(b_1+\beta)t\label{mv1},\\�\nonumber\\
& &m_2(t)=m_{20}+(b_2-\beta)t-m_3(t)\label{mv2},\\�\nonumber\\
& &m_3(t)=m_{30} +m_{3f}(1-e^{-\gamma t}),
\end{eqnarray}
\end{subequations}
where $\beta$ represents the mass transfer between the bodies of masses $m_1$ and $m_2$, the parameters $b_1$  and $b_2$ represent the increases in the effective masses of the bodies $m_1$ and $m_2$ due to the increases of the local particle densities of the spheres in which we define the galaxy masses when using the GADGET-2 code, $m_{30}$  and $m_{3f}$ are the initial and final masses of the third body, and $\gamma$  is related to the rate of mass gain of the third body from the second one.
\subsection{Model-2}
In this model, we assume again that our reference system is located in the body with mass $m_1$, and that the trajectory of the body with mass $m_3$  follows the motion of the body with mass $m_2$ in a fixed specified way. Thus, equations (\ref{e1}) and (\ref{e2}) are not taken into account, and the motion of the system is reduced to the equation
\begin{equation}\label{em21}
\frac{d(m_2{\bf v})}{dt}=-G\frac{m_2m_1}{|{\bf x}|^{3/2}}{\bf x}+G\frac{m_2m_3}{|{\bf x'-x}|^{3/2}}({\bf x'}-{\bf x}),
\end{equation}
where the motion of the body with mass $m_3$ will be determined by the expression
\begin{equation}\label{em22}
{\bf x'}(t)={\bf x}(t-t_r)+\alpha(t-t_r) \widehat{\bf x}(t-t_r),
\end{equation}
where $t_r$ is the retarded time which is the time delay for movement of the third body, after the second body has started to move   (used to simulate the tail leaving the second galaxy). In this model, if $t_r=0$ the second and third bodies would be moving  tangentially on the same radial line, separated by some defined distance, on their tangential motion on the same radial line.  The function  $\alpha(t-t_r)$ is a function which takes into account the separation of the third body from the second one, $\alpha(t-t_r)=\alpha_0(2-e^{-\omega_a(t-t_r)})$.\\ \\  
Equations (\ref{em11}), (\ref{em12}), and (\ref{em21}) are solved numerically, with the functions $m_i(t)$ for $i=1,2,3$ determined qualitatively from the N-body GADGET-2 code, when two galaxies are interacting. The idea is to study the behavior of the separation between the body with mass $m_1$ and the body with mass $m_2$,
\begin{equation}\label{dd}
d(t)=|{\bf x}(t)|,
\end{equation}
and compare  this parameter with the same parameter obtained using the GADGET-2 code.
\section{Simulations of two interacting galaxies with GADGET-2}
The GADGET-2 code is a program that models a galaxy and galaxy interactions. Each galaxy has a bulge, a  disk, central black holes, gas, and dark matter halo. The description and capabilities of this program are given in  \cite{gadget2,gadget2-2,gadget2-3}.
If   $r_{200}$ is the radius containing $75\%$ of the galaxy mass, where the density is 200 times the critical density and $v_{200}$ is the rotation velocity or velocity dispersion at this radius, then the total mass of the galaxy within this radius  is  $M_{200}=v_{200}^2 r_{200}/ G$, where $G$ is the gravitational constant.
\begin{equation}
M_{200}=M_{bulge}+M_{halos}+M_{disk}.
\end{equation}
The halo and bulge densities, as a function of the radius ($r$) of the galaxy, follow the expression 
\begin{equation}
\rho_{halo,bulge}(r)=\frac{M_{halo, bulge}}{2\pi}\frac{a_{h,b}}{r(r+a_{h,b})},
\end{equation}
where the parameter $a_{h,b}$ is given in term of the concentration index $c$ as $a_h=r_s\sqrt{2[\ln(1+c)-c/(1+c)}$, and 
the halo scale length is defined as $r_s=r_{200}/c$ and  the proportion bulge scale length per scale radius $f_b=a_b/R_d$, where $R_d$ is the scale length of the disk. The angular momentum of the halo is given by
\begin{equation}
J=\lambda\sqrt{2GM_{200}^3r_{200}/f_c},
\end{equation}
where $\lambda$ is the twist parameter, and the parameter $f_c$ is written in terms of the concentration index as $f_c=c[1-1/(1+c)^2-2\ln(1+c)/(1-c)]/(2[\ln(1+c)-c/(1+c)]^2)$. The fractional angular momentum of the bulge is defined as $J_b=JM_{bulge}/M_{200}$. The disk (stars) density varies as
\begin{equation}
\rho_{*}(R,z)=\frac{M_{*}}{4\pi z_0 R_d^2}sech^2\left(\frac{z}{2z_0}\right)\exp \left( -\frac{R}{R_d}\right),
\end{equation}
where $R=\sqrt{x^2+y^2}$, $z_0$ is the parameter determining the thickness of the disk. The angular momentum of the disk is $J_d=j_dJ$,  with the free parameter $j_d$. Similarly, the gas in the disk is modelled as
\begin{equation}
\Sigma_{gas}=\frac{M_{gas}}{2\pi h_r^2}\exp \left(-\dfrac{r}{h_r} \right),
\end{equation} 
where $h_r$ is the scale length of the gas profile, and $M_{gas}+M_{*}=M_{disk}$. 
In addition, the vertical structure of gas in asymmetric galaxies is governed by the equation
\begin{equation}
\frac{1}{\rho_{gas}}\frac{\partial P}{\partial z}+\frac{\partial \Phi}{\partial z}=0,
\end{equation} 
where $\Phi$ is the gravitational potential due to the total mass of the gas.\\�\\
Our simulation of two interacting galaxies with the GADGET-2 code were performed with the following set of parameters  for the galaxies $G_1$ and $G_2$ ($N_i$ being the number of elements of each part of the galaxy):
\begin{eqnarray*}
& & {\bf G_1}:\\
& & c=13.0,\quad v_{200}=220~\mathrm{km/s},\quad \lambda=0.033,\quad M_{disk}/M_{200}=0.0,\quad M_{bulge}/M_{200}=0.007,\quad a_b/R_d=1.0, \\�
& & N_{halo}=150,000,\quad N_{disk}=0.0,\quad N_{gas}=0.0,\quad N_{bulge}=50,000,
\end{eqnarray*}\\ where $R_d=3.7397$ h$^{-1}$kpc.
\begin{eqnarray*}
& & {\bf G_2}:\\
& & c=5.0,\quad v_{200}=100~\mathrm{km/s},\quad \lambda=0.033,\quad M_{disk}/M_{200}=0.041,\quad M_{bulge}/M_{200}=0.01367,\quad M_{gas}/M_{disk}=0.1\\�
& & a_b/R_d=z_0/R_d=0.2,\quad J_{disk}/J=0.041, \\�
& & N_{halo}=300,000,\quad N_{disk}=200,000,\quad N_{gas}=20,000,\quad N_{bulge}=10,000,
 \end{eqnarray*}where $R_d=1.978$ h$^{-1}$kpc.\\
\\
 Figure 1 shows the initial configuration of both galaxies with the regions (circles) representing spheres used to define the total masses of the galaxies  in the planes x-y and x-z. The mass variation depends on the size of these circles. Figure 2 and 3 show the mass evolution of both galaxies during their interaction for several regions of different sizes  (as the size increases, the number of elements increases and the mass increases, as shown initially in both figures) 
for the galaxies of mass $m_1$ and $m_2$, where the size of the regions are $ 50~$h$^{-1}$kpc and $50/2.2~$h$^{-1}$kpc  longer than the  previous one, for G1 and G2 ($ h$ is the normalization of the Hubble constant $H_0$,  $h= H_0/100$ km s$^{-1}$ Mpc$^{-1}$ and observationally, $h \sim  0.7$, and 1 kpc $ \approx 206264806~$au $\approx 149.6\times10^6   ~$km)
The maxima on the mass of the second galaxy, shown in Figure 3, are due to excess or counting the elements when the two galaxies are  colliding with each other,  and the minima correspond when they are separated enough that this counting is
well defined.  Figure 4, green solid line,  shows the  evolution of the separation 
 of the centre of the bulges of the two galaxies as a function of time, where we can see the shrinking damping oscillations of the system. To make the calculation of the distance between these two galaxies, we first located the centres of both galaxies (the points of highest density) and separate these centres a distance $d(0)=120~$h$^{-1}$kpc in the x-direction. Then, we follow the evolution of  the separation. This spatial scale covers the typical separations and spatial scales of filaments seen in interactiing galaxies like NGC 4038/4039 \cite{lahen} and merger remnants like NGC 7252 \cite{hibbard}.\\
\begin{figure}[H]
\includegraphics[width=0.7\textwidth]{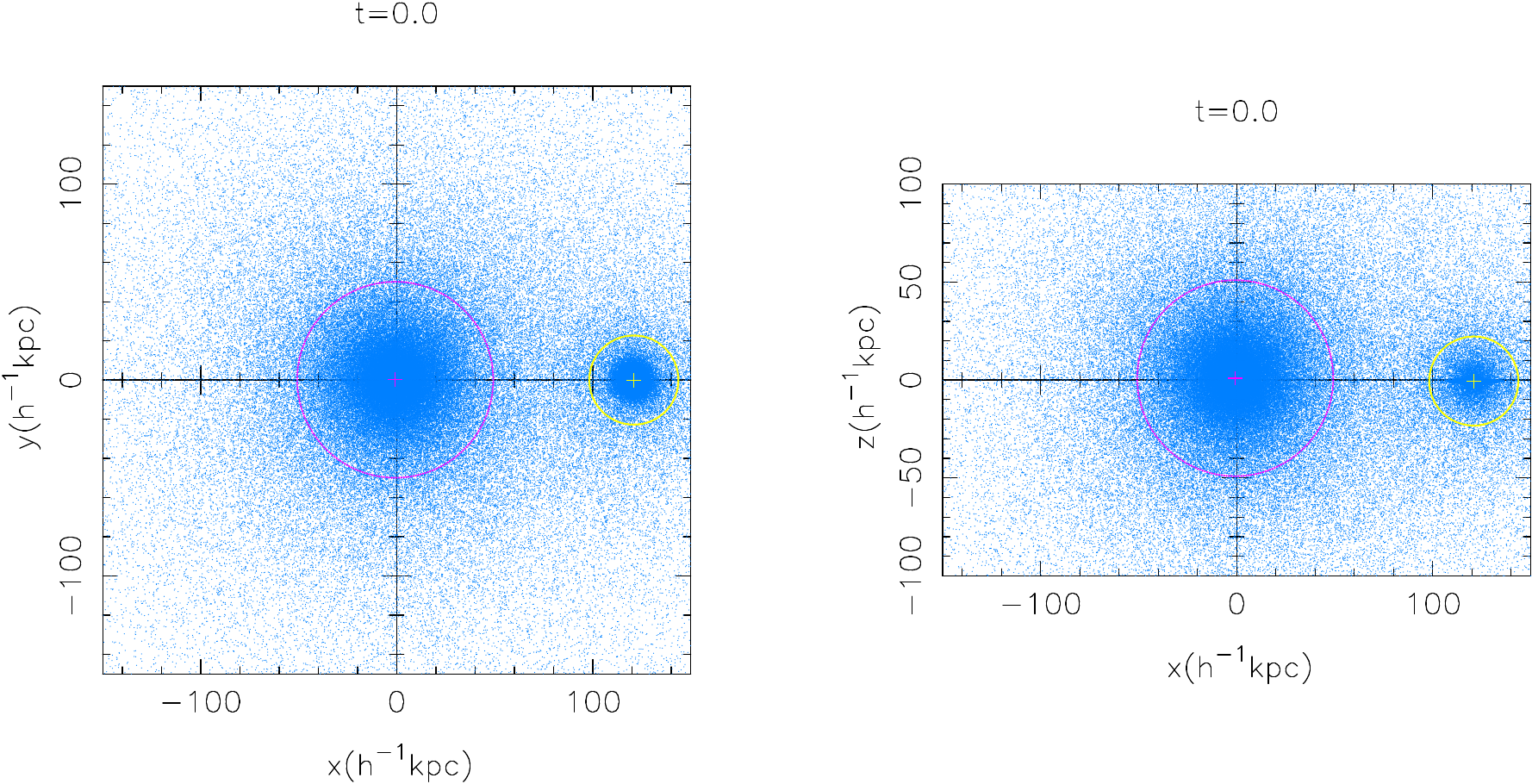}
\centering
    \caption{Initial configuration of both galaxies and the circles used to calculate total masses with the GADGET-2 code.}
\end{figure}
\newpage
\begin{figure}[H]
\includegraphics[width=\textwidth]{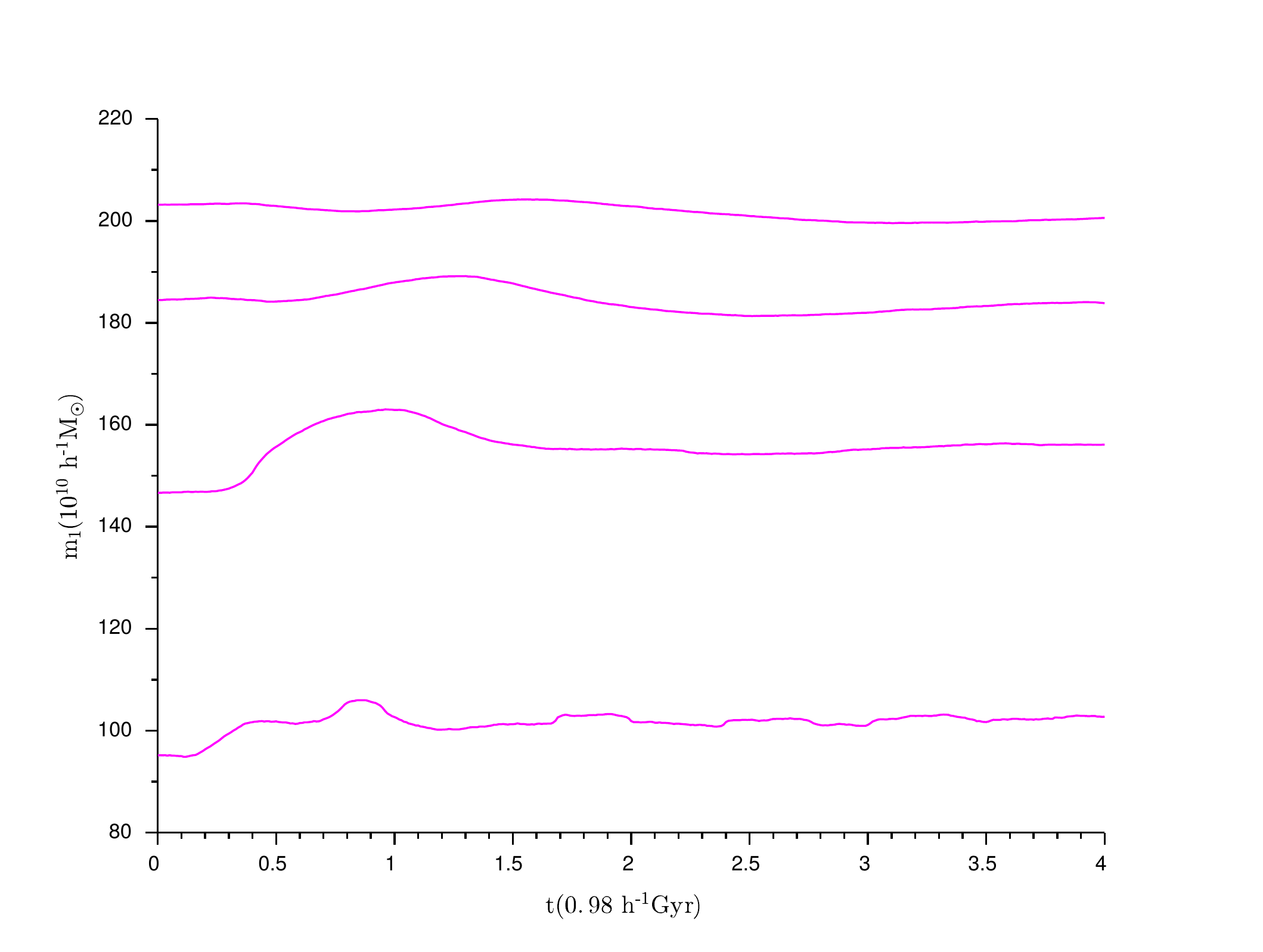}
\centering
    \caption{Mass variation of Galaxy 1, for spheres of different sizes.}
\end{figure}
\begin{figure}[H]
\includegraphics[width=\textwidth]{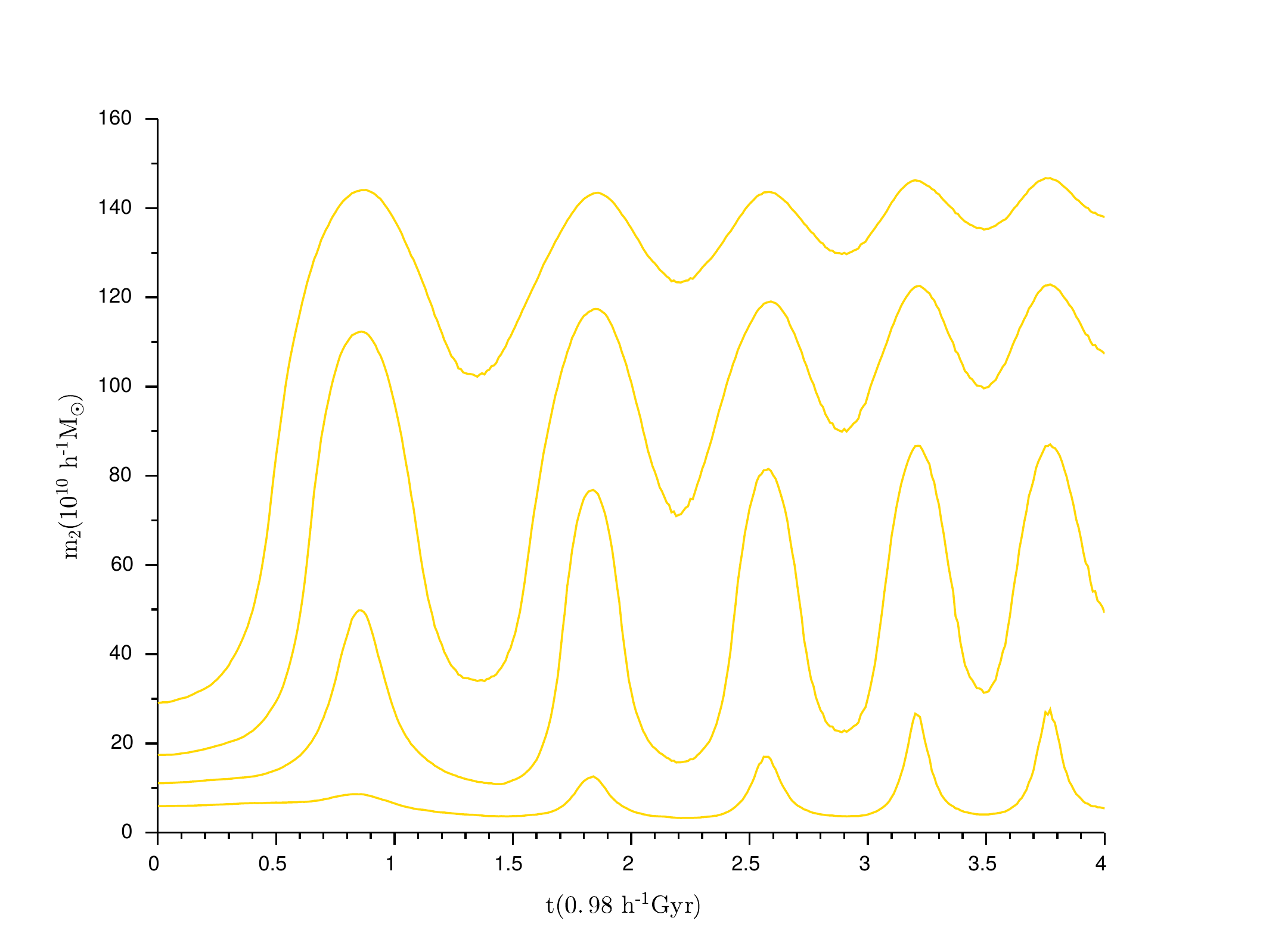}
\centering
    \caption{Mass variation of Galaxy 2, for spheres  of  different sizes.}
\end{figure}
\newpage
\section{ Analytical result and comparison}
As we can see from Figures 1,2 and 3, we can consider the mass of the three bodies and the values of the parameters for our analytical Model-1 (without retarded time) as:\\
\begin{eqnarray*}
& & 
 m_{10}=   185.513000\cdot 10^{10}\ \mathrm{h}^{-1} \mathrm{M}_\odot,\quad
 m_{20}=   17.4265690\cdot 10^{10}\ \mathrm{h}^{-1} \mathrm{M}_\odot,\quad
 m_{30}=m_{3f}=   1.74265695\cdot 10^{10}\ \mathrm{h}^{-1} \mathrm{M}_\odot ,\\
& &  
 \beta= 1.30699262 \ \mathrm{f}^{-1}\mathrm{M}_\odot \mathrm{year}^{-1},\quad
 b_1=   74.2051983 \ \mathrm{f}^{-1}\mathrm{M}_\odot \mathrm{year}^{-1},\quad
 b_2=   21.7832112 \ \mathrm{f}^{-1}\mathrm{M}_\odot \mathrm{year}^{-1},\\
& & 
 \gamma=   1.0 \ \mathrm{f}^{-1}\mathrm{h{Gyear}}^{-1},\quad
 t_r=   0.0\ \mathrm{f}^{-1}\mathrm{tGyear},\quad
 \vert{\bf x}_0-{\bf x'}_0\vert =   24.3476219   \  \mathrm{h}^{-1} \mathrm{kpc},\\
& &  
 x_{0}=   121.73811340332031\  \mathrm{h}^{-1} \mathrm{kpc},\quad
 y_{0}=   0.0\  \mathrm{h}^{-1} \mathrm{kpc},\quad
 v_{x}(0)=   0.0\ \mathrm{km/s},\\
& & 
 v_{y}(0)=   181.02100930177266 \ \mathrm{km/s},\quad
 v'_{x}(0)=   166.4443359375\  \mathrm{km/s},\quad
 v'_{y}(0)=   221.70454793214924\   \mathrm{km/s},
 \end{eqnarray*}\\
where M$_\odot$ represents the Solar Mass (M$_\odot=1.989\times 10^{30}$kg) and the dimensionless parameter $\mathrm{f} = 0.98$. For this same model but with retarded time ($t_r$), the parameters are:\\
\begin{eqnarray*}
& & 
 m_{10}=   185.513000\cdot 10^{10}\ \mathrm{h}^{-1} \mathrm{M}_\odot,\quad
 m_{20}=   17.4265690\cdot 10^{10}\ \mathrm{h}^{-1} \mathrm{M}_\odot,\quad
 m_{30}=m_{3f}=   1.74265695\cdot 10^{10}\ \mathrm{h}^{-1} \mathrm{M}_\odot ,\\
& &  
 \beta= 0.430285633 \ \mathrm{f}^{-1}\mathrm{M}_\odot \mathrm{year}^{-1},\quad
 b_1= 137.417049    \ \mathrm{f}^{-1}\mathrm{M}_\odot \mathrm{year}^{-1},\quad
 b_2= 21.5142822    \ \mathrm{f}^{-1}\mathrm{M}_\odot \mathrm{year}^{-1},\\
& & 
 \gamma=0.987654328  \ \mathrm{f}^{-1}\mathrm{h{Gyear}}^{-1},\quad
 t_r=   0.05 \ \mathrm{h}^{-1}\mathrm{tGyear},\quad
 \vert{\bf x}_0-{\bf x'}_0\vert = 24.3476219  \  \mathrm{h}^{-1} \mathrm{kpc},\\
& &  
 x_{0}=121.73811340332031   \  \mathrm{h}^{-1} \mathrm{kpc},\quad
 y_{0}=0.0   \  \mathrm{h}^{-1} \mathrm{kpc},\quad
 v_{x}(0)=0.0  \   \mathrm{km/s},\\
& & 
 v_{y}(0)=181.02100930177266  \   \mathrm{km/s},\quad
 v'_{x}(0)=0.0   \   \mathrm{km/s},\quad
 v'_{y}(0)=388.14888386964924   \   \mathrm{km/s}.\\�
 \end{eqnarray*} 
The parameters for our analytical Model-2 can be chosen as:\\
\begin{eqnarray*}
& & 
 m_{10}=   185.513000\cdot 10^{10}\ \mathrm{h}^{-1} \mathrm{M}_\odot,\quad
 m_{20}=   17.4265690\cdot 10^{10}\ \mathrm{h}^{-1} \mathrm{M}_\odot,\quad
 m_{30}=0, m_{3f}=   1.74265695\cdot 10^{10}\ \mathrm{h}^{-1} \mathrm{M}_\odot ,\\
& &  
 \beta=0.430285633  \ \mathrm{f}^{-1}\mathrm{M}_\odot \mathrm{year}^{-1},\quad
 b_1=4.58056778  \ \mathrm{f}^{-1}\mathrm{M}_\odot \mathrm{year}^{-1},\quad
 b_2=21.5142822   \ \mathrm{f}^{-1}\mathrm{M}_\odot \mathrm{year}^{-1},\\
& & 
 \gamma= 0.987654328 \ \mathrm{f}^{-1}\mathrm{h{Gyear}}^{-1},\quad
 t_r=   0.05 \ \mathrm{h}^{-1}\mathrm{tGyear},\quad
 \alpha_0= 24.3476219 \  \mathrm{h}^{-1} \mathrm{kpc},\\
& & 
 \omega_\alpha = 0.987654328 \ \mathrm{f}^{-1}\mathrm{h{Gyear}}^{-1},\quad  
 x_{0}= 121.73811340332031  \  \mathrm{h}^{-1} \mathrm{kpc},\quad
 y_{0}= 0.0  \  \mathrm{h}^{-1} \mathrm{kpc},\\
& &
 v_{x}(0)=0.0  \   \mathrm{km/s},\quad
 v_{y}(0)= 181.02100930177266 \   \mathrm{km/s}.\\�
 \end{eqnarray*} 
Solving numerally the equations (\ref{em11}), (\ref{em12}), and (\ref{em21}), we obtain  Figure 4, where the parameter ({\ref{dd}) is plotted for the two mathematical models, Model-1 corresponds to the dotted green curve (no retarded time for the third body) and the dotted-dashed yellow curve (with retarded time $t_r=5~$h$^{-1}$Gyr for the third body),  
Model-2 corresponds to the dashed blue curve (with a retarded time for the second body of $t_r=5~$h$^{-1}$Gyr),  and the result obtained with the GADGET-2 code (continuous red curve).  As one can see, there is a qualitatively good agreement of the GADGET-2 code with both models, and Model-2 with Model-1 (with and without retarded time)  have almost the same behavior, even though Model-2 is an oversimplifying approximation  of Model-1. \\
Figure 5 and 6 show the motion of the second and third bodies (relative to the first body) on the plane of  motion, and one can see that the third body tends to scape from the system (simulating the tail of the second Galaxy moving out of the binary system).  Of course, one can not expect similar behavior since in  Model-2 we have forced the third body to follow the second body 
at some retarded time ($ t_r=5~h^{-1}$Gyr ).  The small square mark on Figures 5 and 6 indicates the position of the second body when the third body is starting to move, due to retarded time.
\begin{figure}[H]
\includegraphics[width=\textwidth]{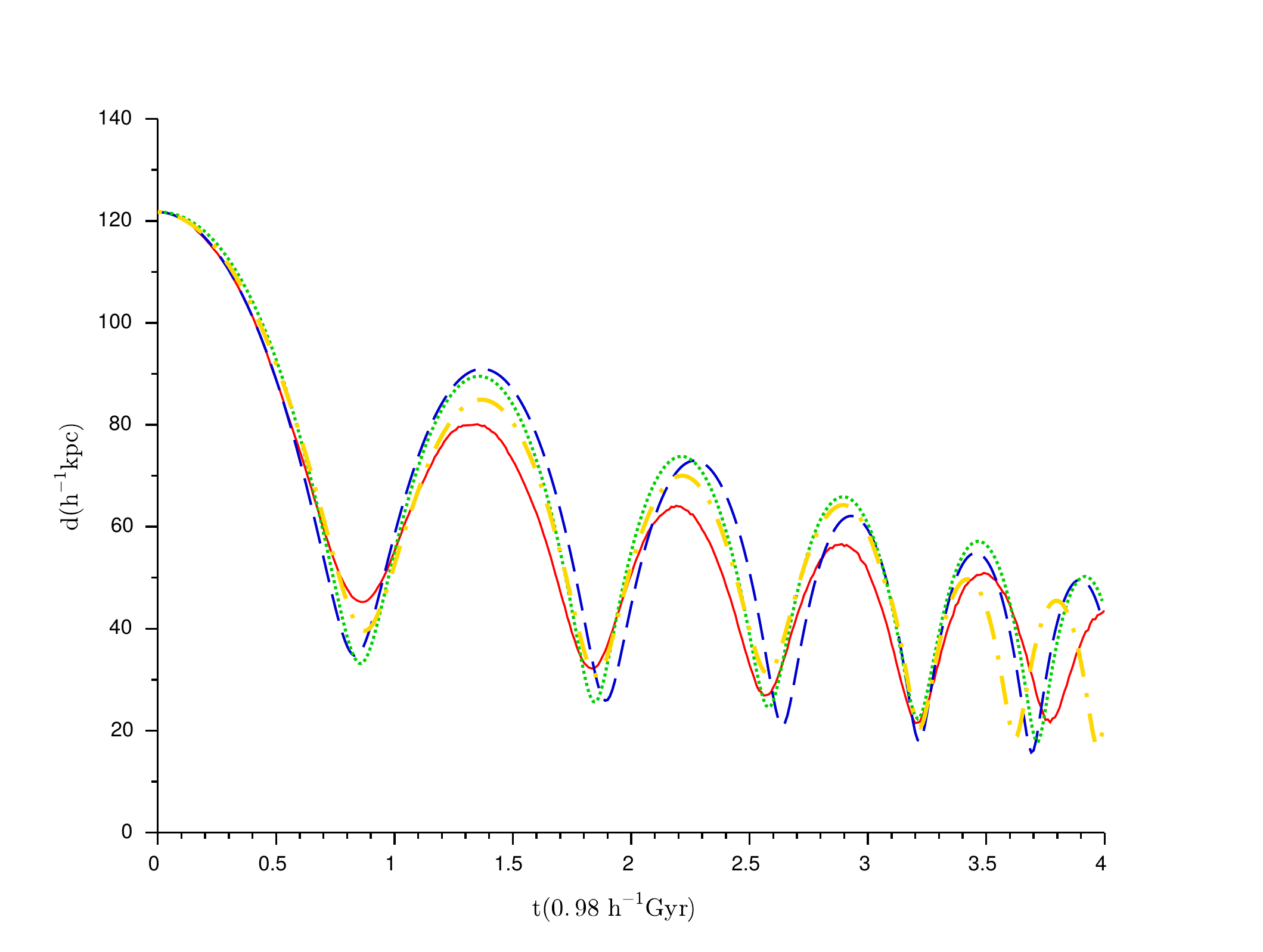}
\centering
    \caption{Distances between the two galaxies as a function of time, GADGET-2 solid red curve, Model-2 dashed blue curve, Model-1 without retarded time is dotted green curve and discontinuous yellow curve with retarded time.}
\end{figure}
\begin{figure}[H]
\includegraphics[width=\textwidth]{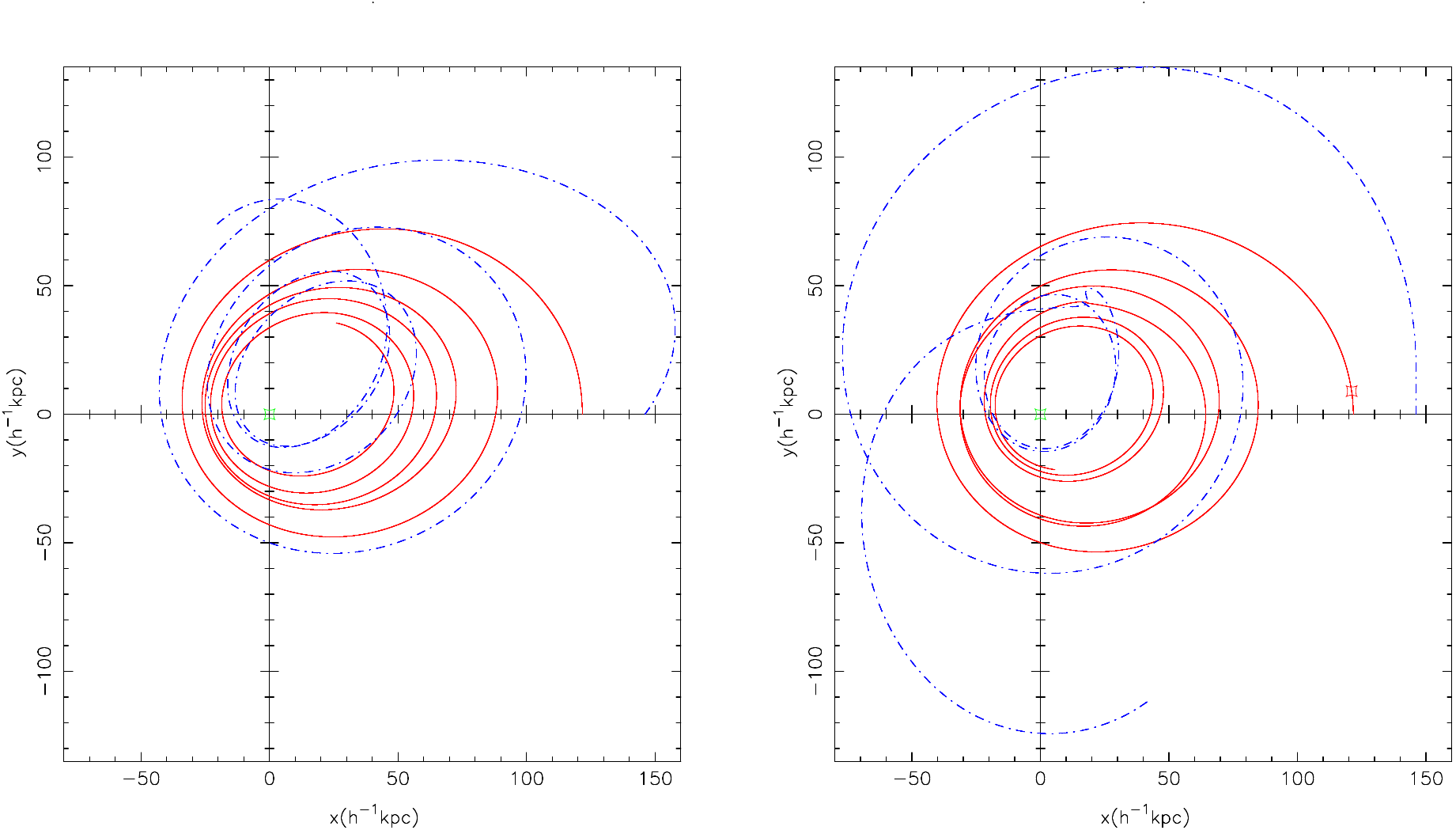}
\centering
    \caption{Model-1: Motion of the two bodies in the plane: Left hand side, without retarded time. Right hand side, with retarded time. Continuous red line represents the second body, and blue dashed-dot line represents the third body.}  
\end{figure}
\begin{figure}[H]
\includegraphics[width=0.6\textwidth]{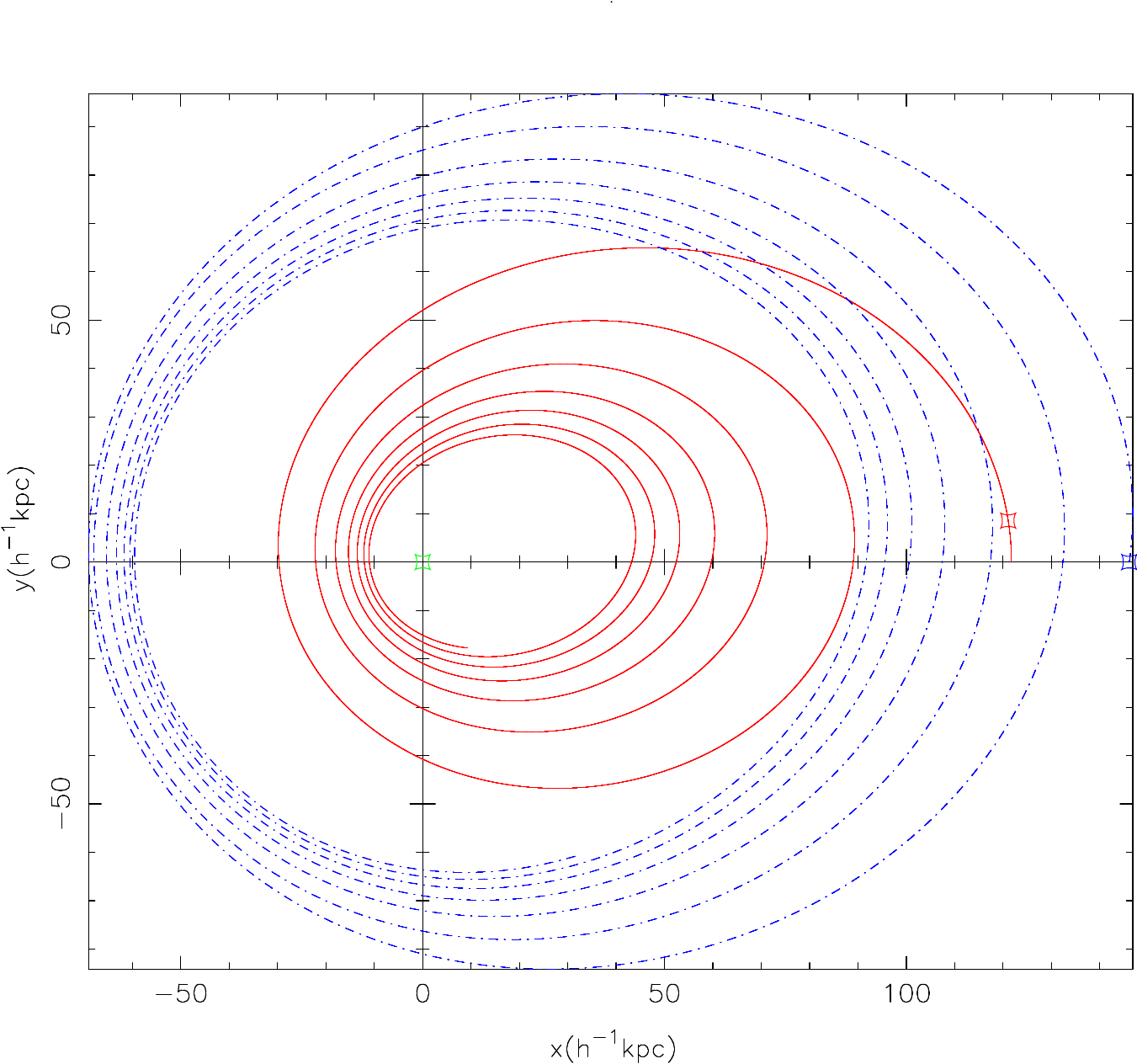}
\centering
    \caption{Model-2: Motion of the two bodies in the plane. Continuous red line represents the second body, and blue dashed-dot line represents the third body.}
\end{figure}
\newpage
\section{Results and Comments}
We have used two mathematical models for the restricted three-body problem of variable masses  to obtain a qualitative  description of two interacting galaxies  with mass exchange and mass loss.
The main point is the use of the third body as a ghost galaxy which simulated the mass lost by the second galaxy in the form of a tail, and this ghost galaxy increases its mass due to the mass lost by the second body. 
 We were interested in predicting the time evolution of the distance separation of the two bulges of the interacting galaxies, using two separate models, one as two bodies with mass variation, the other with one body with mass variation, using  parameters deduced by the use of the  GADGET-2 N-body simulation code for two interacting galaxies.
 \vskip0.5pc\noindent
The general result we have obtained is that  the time dependent  distance between the bulges of the two galaxies, calculated with our mathematical models and  
 deduced from the GADGET-code, agree well qualitatively using both our  mathematical models.  This agreement is 
  better with our Model-1 (three-body problem, without retarded time) than with Model-2, but  both have almost the same behavior. This implies that for a few orbits of two interacting galaxies (before they merge as a single bigger galaxy),  we can use this very simple model to know qualitatively the trajectory of the bulges of the two galaxies. We need to point out that one needs just a single run on N-bodies code to obtain the necessary parameters for the analytical models and to know the dynamics of the budges in much more shorter time than the simulation code. As a final comment, we expect that including also the position depending of the masses of the bodies in our models, the results can be improve  appreciably to match with simulation code.

\newpage
\section*{Acknowledgements}
We want to thank the INAOE institute in Tonantzintla, Puebla, M\'exico, to allow us to have access to their installations and for its financial support. In particular, we want to thank Dr. Ivânio Puerari and   M.C. Diego Valencia Enr\'{\i}quez for their help  teaching to one of us how to use GADGET-2 code.

\end{document}